\documentclass[twocolumn]{jpsj2} 
\usepackage{bm,color}

\title{Mechanism of Lattice-Distortion-Induced Electric-Polarization Flop in the Multiferroic Perovskite Manganites}

\author{Masahito \textsc{Mochizuki}$^{1}$\thanks{E-mail address:
mochizuki@erato-mf.t.u-tokyo.ac.jp} and Nobuo \textsc{Furukawa}$^{2}$}

\inst{$^1$Multiferroics Project, ERATO, Japan Science and Technology Agency (JST),\\
c/o Department of Applied Physics, The University of Tokyo, 7-3-1 Hongo, Bunkyo-ku, Tokyo 113-8656, Japan \\
 $^2$Department of Physics and Mathematics, Aoyama Gakuin University, Sagamihara, Kanagawa 229-8558, Japan}

\abst{We theoretically study a competition between two types of spin cycloids ($ab$-plane and $bc$-plane ones) in the multiferroic perovskite manganites, which is an origin of intriguing magnetoelectric phenomena in these compounds. Analysis of a microscopic model using the Monte-Carlo method reveals that their competition originates from a conflict between the single-ion anisotropy and the Dzyaloshinsky-Moriya interaction. It is demonstrated that the conflict can be controlled by tuning the second-neighbor spin exchanges through the GdFeO$_3$-type distortion, which leads to a cycloidal-plane flop from $ab$ to $bc$ observed in the solid-solution systems like Eu$_{1-x}$Y$_x$MnO$_3$ and Gd$_{1-x}$Tb$_x$MnO$_3$ with increasing $x$.}

\kword{multiferroics, Mn perovskite, phase diagram, Dzyaloshinskii-Moriya interaction}

\begin{document}
\maketitle
Magnetoelectric (ME) coupling in concurrently magnetic and ferroelectric [multiferroic (MF)] materials causes a lot of fascinating ME phenomena~\cite{Kimura03a,ReviewMF}. Intensive experiments revealed magnetic-field ($H$) induced ferroelectric-polarization ($P$) flops~\cite{Kimura05}, giant magnetocapacitance effects~\cite{Kimura05,Goto04,Kagawa09}, and electrically activated magnon excitations (termed electromagnons)~\cite{Smolenski82,Pimenov06a} in the typical MF materials of perovskite manganites $R$MnO$_3$ with $R$ being a rare-earth ion. The coupling and the phenomena have attracted enormous interest from viewpoints of both science and application since their manipulations enable us magnetic control of ferroelectricity and electric control of magnetism. 

The ferroelectricity in $R$MnO$_3$ is induced by the cycloidally ordered spin structure through the inverse Dzyaloshinsky-Moriya (DM) mechanism~\cite{Katsura05,Sergienko06a,Mostovoy06}. The $bc$-plane ($ab$-plane) spin cycloid propagating along $b$ generates $P$ along the $c$ ($a$) axis~\cite{Kenzelmann05}. Since the first report of MF nature in TbMnO$_3$, a number of MF materials of spiral-magnetism origin have been discovered in succession, and these materials have turned out to exhibit similar interesting ME phenomena also. However, up to now, mechanisms of these ME phenomena have not been clarified yet in spite of much effort.

\begin{figure}[tdp]
\includegraphics[scale=1.0]{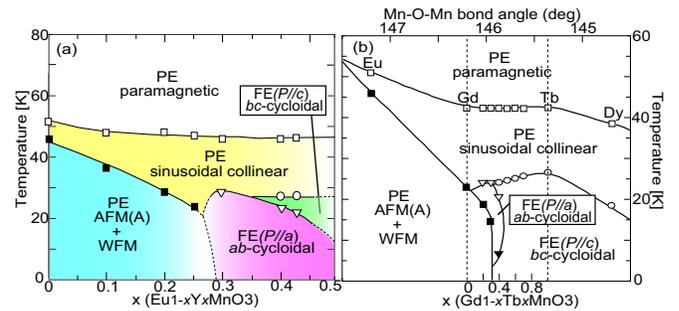}
\caption{(Color online) Experimental $T$-$x$ phase diagrams of (a) Eu$_{1-x}$Y$_x$MnO$_3$ and (b) Gd$_{1-x}$Tb$_x$MnO$_3$ reproduced from Ref.\cite{Yamasaki07b} and Ref.\cite{Goto05}, respectively. PE and FE denote paraelectric and ferroelectric phases, and AFM(A)+WFM denotes the A-type antiferromagnetic phase with weak ferromagnetism.}
\label{Fig01}
\end{figure}
An important clue to approach these issues were recently provided from precise experimental studies on the phase diagrams of $R$MnO$_3$. When we vary the magnitude of GdFeO$_3$-type distortion continuously by using solid solutions like Eu$_{1-x}$Y$_x$MnO$_3$ (EYMO)~\cite{Hemberger07,Yamasaki07b} and Gd$_{1-x}$Tb$_x$MnO$_3$ (GTMO)~\cite{Goto05}, the 90$^{\circ}$ reorientation of $P$ from $P$$\parallel$$a$ to $P$$\parallel$$c$ is observed as Y or Tb concentration $x$ increases, which is accompanied by a cycloidal-plane flop from $ab$ to $bc$~\cite{Yamasaki08} ----- see Fig.~\ref{Fig01}.

Interestingly we notice that the competition between these two types of spin cycloids ($ab$/$bc$ competition) has a universal relevance to the above ME phenomena. The $P$ flop under an applied $H$ is a consequence of the competition controlled by $H$. The theoretically proposed electromagnon is an oscillation of the spin-cycloidal plane\cite{Katsura07}. The giant magnetocapacitance effect was experimentally ascribed to the electric-field-driven motion of domain walls between the $ab$- and $bc$-cycloidal phases~\cite{Kagawa09}.

These indicate that clarification of origin of the $ab$/$bc$ competition in $R$MnO$_3$ is a key to understanding the ME phenomena in these MF materials. Recent theoretical studies reproduced the spiral spin order in $R$MnO$_3$~\cite{Sergienko06a,SDong08}. However the origin of the $ab$/$bc$ competition and a mechanism of the cycloidal-plane flop were not addressed. 

In this Letter, by taking $R$MnO$_3$ as a typical example, we theoretically study the $ab$/$bc$ competition by constructing a microscopic model for the Mn $3d$-spin system. We reveal that the competition originates from a conflict between the single-ion anisotropy and the DM interaction. This competition is controlled by the second-neighbor spin exchanges enhanced by the GdFeO$_3$-type distortion, which leads to a cycloidal-plane flop in EYMO and GTMO with increasing $x$. We demonstrate that their $T$-$x$ phase diagrams are naturally reproduced as its consequence. Our finding provides a fundamental insight on the ME phenomena in the MF materials of magnetic-spiral origins.

\begin{figure}[tdp]
\includegraphics[scale=1.0]{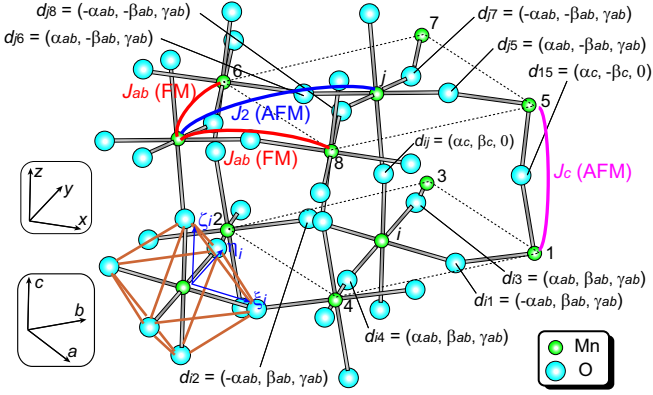}
\caption{(Color online) Superexchange interactions, tilted local axes $\xi_i$, $\eta_i$, and $\zeta_i$ attached to the $i$-th MnO$_6$ octahedron, and Dzyaloshinsky-Moriya vectors in $R$MnO$_3$.}
\label{Fig02}
\end{figure}
We start with a classical Heisenberg model with some additional interactions and magnetic anisotropies on a cubic lattice, in which the Mn $S$=2 spins are treated as classical vectors. The Hamiltonian consists of four terms as
$\mathcal{H}=\mathcal{H}_{\rm ex}+\mathcal{H}_{\rm sia}
+\mathcal{H}_{\rm DM}+\mathcal{H}_{\rm cub}$~\cite{Matsumoto70}.
We adopt meV as the unit of energy.

The first term $\mathcal{H}_{\rm ex}$ denotes superexchange interactions; ferromagnetic (FM) $J_{ab}$ on the in-plane nearest-neighbor bonds, antiferromagnetic (AFM) $J_2$ on the in-plane second-neighbor bonds along the $b$ axis, and AFM $J_c$ on the bonds along the $c$ axis as shown in Fig.~\ref{Fig02}. The second-neighbor exchanges $J_2$ are caused by finite overlaps of $e_g$ orbitals in the $b$ direction mediated by two oxygens.

The second term $\mathcal{H}_{\rm sia}$ denotes the single-ion anisotropy (SIA), which can be written as $\mathcal{H}_{\rm sia}=
D\sum_{i} (\bm S_i \cdot \bm {\zeta}_i/|\bm {\zeta}_i|)^2
+E\sum_{i} (-1)^{i_x+i_y}
[(\bm S_i \cdot \bm {\xi}_i/|\bm {\xi}_i|)^2
-(\bm S_i \cdot \bm {\eta}_i/|\bm {\eta}_i|)^2]$.
Here $\bm \xi_i$, $\bm \eta_i$ and $\bm \zeta_i$ are directional vectors of the tilted local axes attached to $i$-th MnO$_6$ octahedron as shown in Fig.~\ref{Fig02}. Because of this term, the $c$ axis in $R$MnO$_3$ becomes a hard magnetization axis. For the vectors, we use the structural data of EuMnO$_3$~\cite{Dabrowski05}. We have confirmed that the results are not significantly changed even if we use structure data of other $R$MnO$_3$.

We have microscopically determined all the parameters in this model. The spin exchanges $J_{ab}$ and $J_c$ and the SIA parameters $D$ and $E$ have been calculated by using formulae given in Ref.\cite{Gontchar01} and Ref.\cite{Matsumoto70}, respectively. We have found that the values are nearly independent of $R$ as far as vicinities of the multiferroic phases are concerned. We take $J_{ab}$=0.80, $J_c$=1.25, $D$=0.25, and $E$=0.30.

We study the $x$ dependence by varying the value of $J_2$. We consider that main roles of the GdFeO$_3$-type distortion controlled by $x$ on the magnetic properties in $R$MnO$_3$ are induction and enhancement of the second-neighbor antiferromagnetic exchanges $J_2$~\cite{Kimura03b}. This is because their exchange paths contain two oxygen $2p$ orbitals, and the distortion enhances the hybridization between these two $2p$ orbitals. Consequently, the spiral rotation angle $\phi$=180$^{\circ}$$\times$$q_{\rm Mn}$ with $q_{\rm Mn}$ being a spiral wave number increases as $x$ increases (note that a relation $\cos\phi=J_1/(2J_2)$ holds in the simple $J_1$-$J_2$ model). The crucial importance of the second-neighbor exchanges $J_2$ is peculiar in these Mn$^{3+}$ compounds. Because of the $t_{2g}^3e_g^1$ electron configuration, the nearest-neighbor FM coupling $J_{ab}$ is strongly reduced due to the cancellation of opposite contributions from two different orbital sectors, i.e. FM coupling between $S$=1/2 $e_g$ spins and AFM coupling between $S$=3/2 $t_{2g}$ spins. This leads to the increasing importance of the exchanges $J_2$ with cooperative AFM contributions from these two orbital sectors.

The third term $\mathcal{H}_{\rm DM}=\sum_{<i,j>}\bm d_{ij}\cdot(\bm S_i \times \bm S_j)$ denotes the DM interactions~\cite{Dzyaloshinsky58,Moriya60a}. The DM vectors $\bm d_{ij}$ are defined on the Mn($i$)-O-Mn($j$) bonds, and are expressed in terms of five DM parameters, $\alpha_{ab}$, $\beta_{ab}$, $\gamma_{ab}$, $\alpha_c$ and $\beta_c$, as shown in Fig.~\ref{Fig02} reflecting the crystal symmetry. We can deduce their values from results of the first-principles calculation~\cite{Solovyev96} and the electron-spin resonance (ESR) experiments~\cite{Deisenhofer02}. We take $\alpha_{ab}$=0.10, $\beta_{ab}$=0.10, $\gamma_{ab}$=0.14, $\alpha_c$=0.30, and $\beta_c$=0.30. We also study the case for a slightly larger $\alpha_c$ of 0.38 without changing the other parameters.
The last term $\mathcal{H}_{\rm cub}=\frac{a}{S(S+1)}\sum_{i}(S_{xi}^4+S_{yi}^4+S_{zi}^4)$ represents the cubic anisotropy.
The constant $a$ for Mn$^{3+}$ ion in the O$_6$ octahedron was evaluated to be 0.0162 in the ESR experiment~\cite{Gerristen63}.

We calculate thermodynamic properties of this model by using the Monte-Carlo (MC) method. To overcome the slow MC dynamics at low temperature ($T$), we employ the replica exchange MC method~\cite{Hukushima96}. Each exchange sampling is taken after 400 standard MC steps. Typically, we perform 600 exchanges for a system with 48$\times$48$\times$6 sites under the periodic boundary condition. We confirm that the finite-size effect is small enough and never affects our conclusion qualitatively. We calculate specific heat $C_s(T)$ and total spin-helicity vector $\bm h^b(T)$ to determine the transition points and the magnetic structures, which are respectively calculated as
$C_s(T)=\frac{1}{N} \partial \langle \mathcal{H}\rangle / \partial (k_{\rm B}T)$, and $\bm h^b(T)=\frac{1}{N} \langle |\sum_{i} \bm S_i \times \bm S_{i+b} |\rangle/S^2$, where the brackets denote thermal averages. The assignments of transitions and phases are confirmed by the calculated spin-correlation functions.

\begin{figure}[tdp]
\includegraphics[scale=1.0]{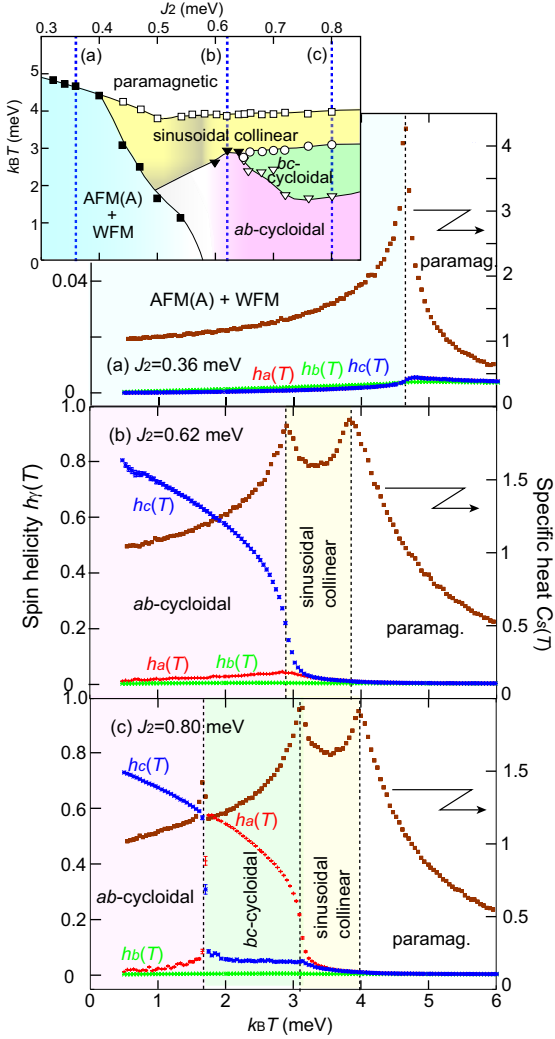}
\caption{(Color online) Calculated temperature profiles of specific heat $C_s(T)$ and total spin helicity $\bm h^b(T)=[h_a(T), h_b(T), h_c(T)]$ for (a) $J_2$=0.36, (b) $J_2$=0.62, and (c) $J_2$=0.80 when $\alpha_c$=0.30.}
\label{Fig03}
\end{figure}
In Figs.~\ref{Fig03}(a)-\ref{Fig03}(c), we show the calculated $C_s(T)$ and $\bm h^b(T)$ when $\alpha_c$=0.30 for various values of $J_2$. Here $h_\gamma(T)$ ($\gamma$=$a$, $b$, $c$) denotes the $\gamma$ component of $\bm h^b(T)$. In the $ab$-cycloidal [$bc$-cycloidal] phase, $h_c(T)$ [$h_a(T)$] has a large value, while other two components are strongly suppressed. On the other hand, in the AFM(A) and sinusoidal collinear phases, all of the three components should be nearly equal to zero.

In Fig.~\ref{Fig03}(a), we can see a single phase transition in $C_s(T)$ for a small value of $J_2$=0.36, which corresponds to a transition from paramagnetic to canted AFM(A) phases. Through this transition, $\bm h^b(T)$ is nearly equal to zero constantly.

For a larger value of $J_2$=0.62, $C_s(T)$ shows two peaks indicative of two thermal transitions ----- see Fig.~\ref{Fig03}(b). $\bm h^b(T)$ is approximately zero constantly through the first transition, at which the system enters into the sinusoidal collinear phase from the paramagnetic phase. Contrastingly, at the subsequent transition into the $ab$-cycloidal phase, its $c$ component $h_c(T)$ starts increasing.

For a further increased value of $J_2$=0.80, the system undergoes three thermal transitions from high to low temperatures ----- see Fig.~\ref{Fig03}(c). The first one is a transition from paramagnetic to sinusoidal collinear phases through which all of the three components of $\bm h^b(T)$ are approximately zero constantly . At the second transition into the $bc$-cycloidal phase, its $a$ component $h_a(T)$ increases, while other two components remain to be small. With further lowering $T$, the $a$ component $h_a(T)$ suddenly drops, while the $c$ component $h_c(T)$ steeply increases at the third transition accompanied by a cycloidal-plane flop from $bc$ to $ab$.

\begin{figure}[tdp]
\includegraphics[scale=1.0]{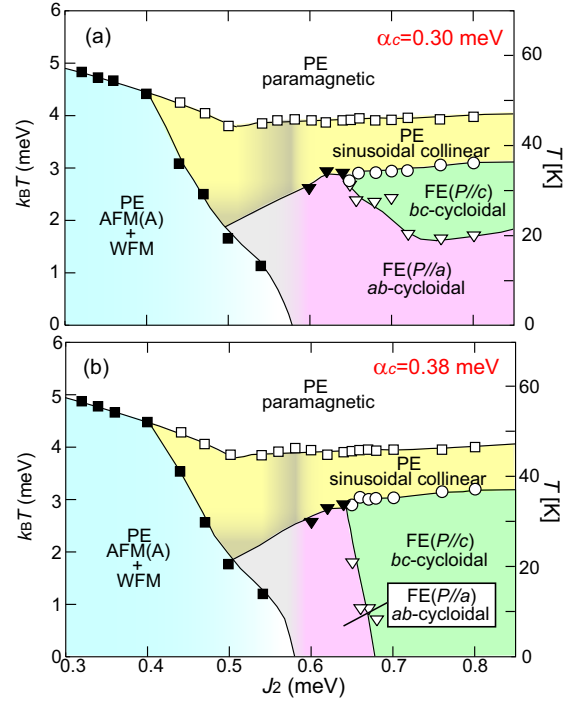}
\caption{(Color online) Theoretical $T$-$J_2$ phase diagrams for (a) $\alpha_c$=0.30 and (b) $\alpha_c$=0.38. PE and FE denote paraelectric and ferroelectric phases expected in the inverse DM model~\cite{Katsura05,Sergienko06a,Mostovoy06}, respectively.}
\label{Fig04}
\end{figure}
In Figs.~\ref{Fig04}(a) and \ref{Fig04}(b), we display theoretically obtained $T$-$J_2$ diagrams for $\alpha_c$=0.30 and $\alpha_c$=0.38, respectively. They are in good agreement with the experimental $T$-$x$ diagrams of EYMO and GTMO shown in Fig.~\ref{Fig01}. For both EYMO with $x$=0.4 and TbMnO$_3$, the value of $J_2$ is calculated to be approximately 0.65 from experimentally obtained $q_{\rm Mn}$~\cite{Yamasaki07b,Goto05}. 
Both diagrams show that the $bc$-cycloidal regime increases with increasing $J_2$, resulting in the cycloidal-plane flop in certain $T$ ranges.

\begin{figure}[tdp]
\includegraphics[scale=1.0]{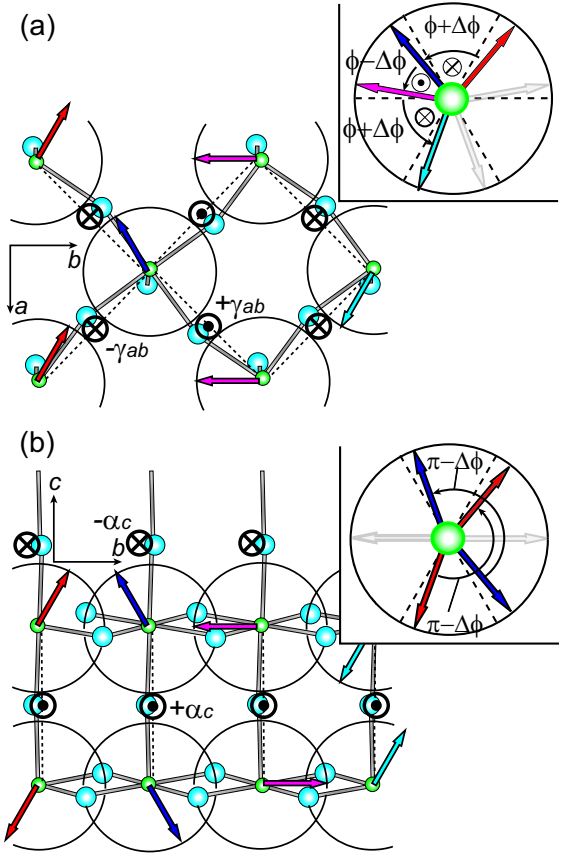}
\caption{(Color online) (a) [(b)] Spin structure in the $ab$-cycloidal [$bc$-cycloidal] state and arrangement of the $c$ [$a$] components of DM vectors on the in-plane [out-of-plane] Mn-O-Mn bonds. The symbols $\odot$ and $\otimes$ express their signs, i.e. $\odot$ for the positive sign and $\otimes$ for the negative sign. Inset shows spin directions with modulated rotation angles due to the DM interactions. Dashed lines indicate spin directions without angle modulations.}
\label{Fig05}
\end{figure}

Here we argue the mechanism of the flop.
In the $ab$-cycloidal state the rotating spins couple dominantly to the $c$ components of DM vectors on the in-plane Mn-O-Mn bonds. Their magnitudes are all equal to $\gamma_{ab}$, and their signs (i.e. $+\gamma_{ab}$ and $-\gamma_{ab}$) are alternately arranged along the $x$ and $y$ bonds ----- see Fig.~\ref{Fig05}(a). Without DM interaction, the spins rotate with the uniform rotation angles of $\phi_{ab}$. On the other hand, in the presence of DM interactions, the rotation angles become to be alternately modulated into $\phi_{ab}+\Delta \phi_{ab}$ and $\phi_{ab}-\Delta \phi_{ab}$ with $\Delta \phi_{ab}>0$ to get an energy gain from the DM interactions ----- see the inset of Fig.~\ref{Fig05}(a). We can derive the energy gain due to this angle modulation as
\begin{eqnarray}
\Delta E_{\rm DM}^{ab}/N
&=&-\gamma_{ab}S^2|\sin(\phi_{ab}-\Delta\phi_{ab})-\sin\phi_{ab}| \nonumber \\
&=&-\gamma_{ab}S^2 |\cos\phi_{ab}| \Delta\phi_{ab}.
\end{eqnarray}
This expression implies that the energy gain $|\Delta E_{\rm DM}^{ab}|$ is reduced with increasing $\phi_{ab}$ because the prefactor $|\cos\phi_{ab}|$ becomes maximum (=1) for $\phi_{ab}$~=0 but decreases as $\phi_{ab}$ increases. Since the angle $\phi_{ab}$ increases as $J_2$ increases, the $ab$-cycloidal state is destabilized with increasing $J_2$ or with increasing GdFeO$_3$-type distortion.

On the other hand, the spins in the $bc$-cycloidal state dominantly couple to the $a$ components of DM vectors on the out-of-plane Mn-O-Mn bonds. Their magnitudes are all equal to $\alpha_c$, and their signs are the same within a plane, but alternate along the $c$ axis ----- see Fig.~\ref{Fig05}(b). Without DM interaction, the angles between adjacent two spins along the $c$ axis are uniformly $\phi_c=\pi$ because of the strong AFM coupling $J_c$. In the presence of DM interactions, the angles again suffer from modulations into $\pi+\Delta\phi_c$ and $\pi-\Delta\phi_c$ with $\Delta \phi_c>0$ ----- see the inset of Fig.~\ref{Fig05}(b). Similar to the $ab$-cycloidal case, we can derive the energy gain due to the angle modulation as
\begin{equation}
\Delta E_{\rm DM}^{bc}/N
=-\alpha_c S^2 \Delta \phi_c,
\end{equation}
irrespective of the value of $J_2$.

As a result, the energetical advantage of $bc$-cycloidal spin state relative to the $ab$-cycloidal one due to the DM interactions, $|\Delta E_{\rm DM}^{bc}-\Delta E_{\rm DM}^{ab}|$, increases as $J_2$ increases. The $bc$-cycloidal spin state is stabilized when the above energy difference dominates over the energetical disadvantage due to the hard-magnetization $c$ axis. The above discussion tells us that the $a$ components of DM vectors on the out-of-plane bonds are relevant to the stability of $bc$-cycloidal spin state. As seen in Figs.~\ref{Fig04}(a) and \ref{Fig04}(b), the diagram for $\alpha_c$=0.38 indeed has a larger regime of $bc$-cycloidal phase than the diagram for $\alpha_c$=0.30.

The diagram of GTMO is reproduced when $\alpha_c$=0.38, and this value is slightly larger than $\alpha_c$=0.30, for which the diagram of EYMO is reproduced. This difference may be due to different $R$-site-radius dependence of the GdFeO$_3$-type distortion. According to Ref.\cite{Hemberger07}, the lattice parameters of TbMnO$_3$ are equivalent to those of EYMO with $x\sim0.85$ although TbMnO$_3$ is expected to be located at $x\sim$~0.4 in the $T$-$x$ diagram in terms of the averaged $R$-site radius. This indicates that the lattice of TbMnO$_3$ is more significantly distorted than expected from comparison to the EYMO system ----- we may have to consider not only the average of $R$-site radii but also their variance~\cite{Tomioka04}. The out-of-plane Mn-O-Mn bond angles in GTMO tend to more deviate from 180$^{\circ}$ than those in EYMO, resulting in a larger value of $\alpha_c$. 

There is a slight difference between the experimental diagram of GTMO and theoretical diagram for $\alpha_c$=0.38 ----- compare Fig.~\ref{Fig01}(b) and Fig.~\ref{Fig04}(b). In the experimental one, the phase boundary between the two cycloidal phases slightly bends, and in a narrow region, the system exhibits a reentrant behavior with successive transitions from $bc$- to $ab$- and again to $bc$-cycloidal phases with lowering $T$. In addition, the $ab$-cycloidal spin order is absent in the ground state. These points are not reproduced in our calculation. This discrepancy may be solved by considering effects of $f$-moments on the $R$ ions, which order below $\sim$10 K. By contrast, in the case of EYMO without interference from $f$-moments, the agreement between the experiment and the calculation is quite good.

In $R$MnO$_3$, the $ab$- and $bc$-cycloidal spin states are stabilized by SIA or DM interaction. Namely, the hard magnetization $c$ axis due to SIA gives a relative stability to the $ab$-cycloidal state, while the $bc$-cycloidal state is stabilized by the DM vectors on the out-of-plane bonds. On the other hand, the $ac$-plane spin cycloid is unfavorable. Recent first-principles calculations for TbMnO$_3$ also confirmed this tendency~\cite{HJXiang08,Malashevich08}.

To summarize, we have studied the spin-cycloidal plane flop from $ab$ to $bc$ in $R$MnO$_3$ by constructing a microscopic model. We have used realistic parameters determined experimentally. We have revealed that the flop occurs due to the competition between SIA and DM interaction controlled by the GdFeO$_3$-type distortion. It has been demonstrated that consideration of this competition naturally reproduces the experimental $T$-$x$ diagrams of EYMO and GTMO, and that incorporation of SIA and DM interaction is indispensable to describe the ME system in $R$MnO$_3$. Note that we have also successfully reproduced the sinusoidal collinear spin phase in the intermediate $T$ regime, whereas previous theories failed~\cite{SDong08}. Our finding provides a useful insight for studying origins and mechanisms of intriguing ME phenomena in the MF materials of magnetic-spiral origin.

We acknowledge valuable discussions with Y. Tokura and N. Nagaosa. MM acknowledge discussions with F. Kagawa, S. Miyahara, N. Kida, H. Murakawa, Y. Yamasaki, and I. Solovyev, and technical advices on the numerical simulations from Y. Motome and D. Tahara.

\end{document}